\documentclass[aps,prc,superscriptaddress,showpacs,nofootinbib,floatfix,twocolumn]{revtex4}

\usepackage{graphicx}	
\usepackage{subfigure}

\newcommand{\dau}{{\rm d+Au}}

\newcommand{\ncol}{{N_{\rm coll}}}
\newcommand{\rda}{{R_{\rm dAu}}}

\begin{document}

\title{Erratum: Cold Nuclear Matter Effects on $J/\psi$ Production as Constrained
by Deuteron-Gold Measurements at $\sqrt{s_{NN}}=200$~GeV
\break Phys. Rev. C77, 024912 (2008)}


\author{A.~Adare}		
\author{S.S.~Adler}		
\author{S.~Afanasiev}		
\author{C.~Aidala}		
\author{N.N.~Ajitanand}		
\author{Y.~Akiba}		
\author{H.~Al-Bataineh}		
\author{J.~Alexander}		
\author{A.~Al-Jamel}		
\author{K.~Aoki}		
% ...
\collaboration{PHENIX Collaboration} \noaffiliation

\date{\today}

\pacs{25.75.Dw} 


\maketitle

All of the experimental data points presented in the 
original paper are correct and unchanged (including 
statistical and systematic uncertainties).  However, herein 
we correct a comparison between the experimental data and a 
theoretical picture using a set of shadowing models for the 
nuclear parton distribution functions (PDFs) combined with a 
nuclear breakup cross section ($\sigma_{\rm breakup}$).  
Under the assumption that a given modified nuclear PDF is 
correct, we put constraints on the $\sigma_{\rm breakup}$ values 
as presented in the original paper in Table V, Section VI.  
In the code that calculated these 
constrained $\sigma_{\rm breakup}$ values we discovered a 
mistake, which we now correct.

Table~\ref{tab:sigma_breakup} shows the new results, 
for which all of the most probable $\sigma_{\rm breakup}$ 
values differ by less than 0.4 mb from those originally 
presented.  However, the one standard deviation 
uncertainties (that include contributions from both the 
statistical and systematic uncertainties on the 
experimental data points) are approximately 30-60\% larger 
than originally reported.

\begin{figure}[h]
\includegraphics[width=1.0\linewidth]{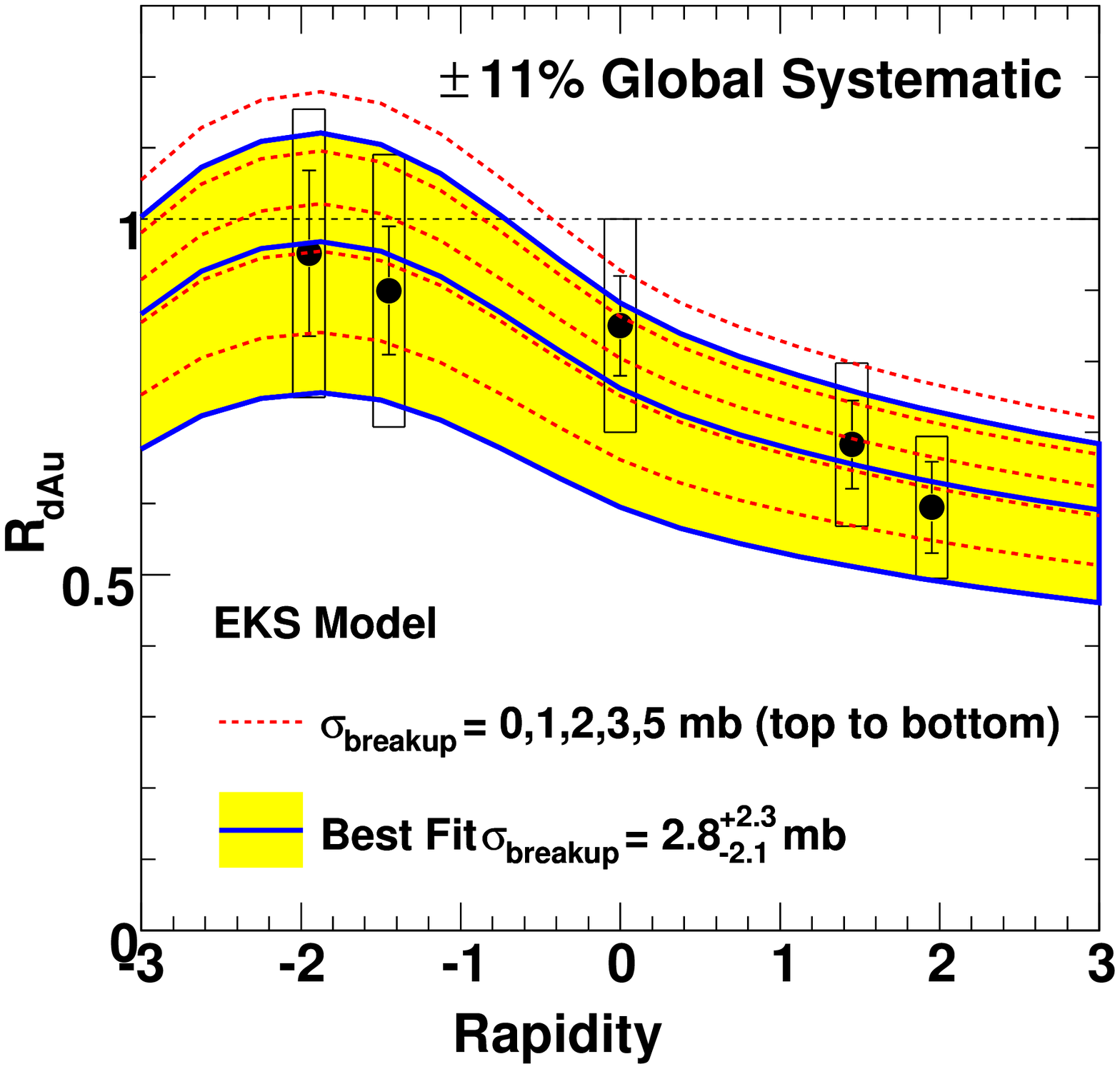}
\includegraphics[width=1.0\linewidth]{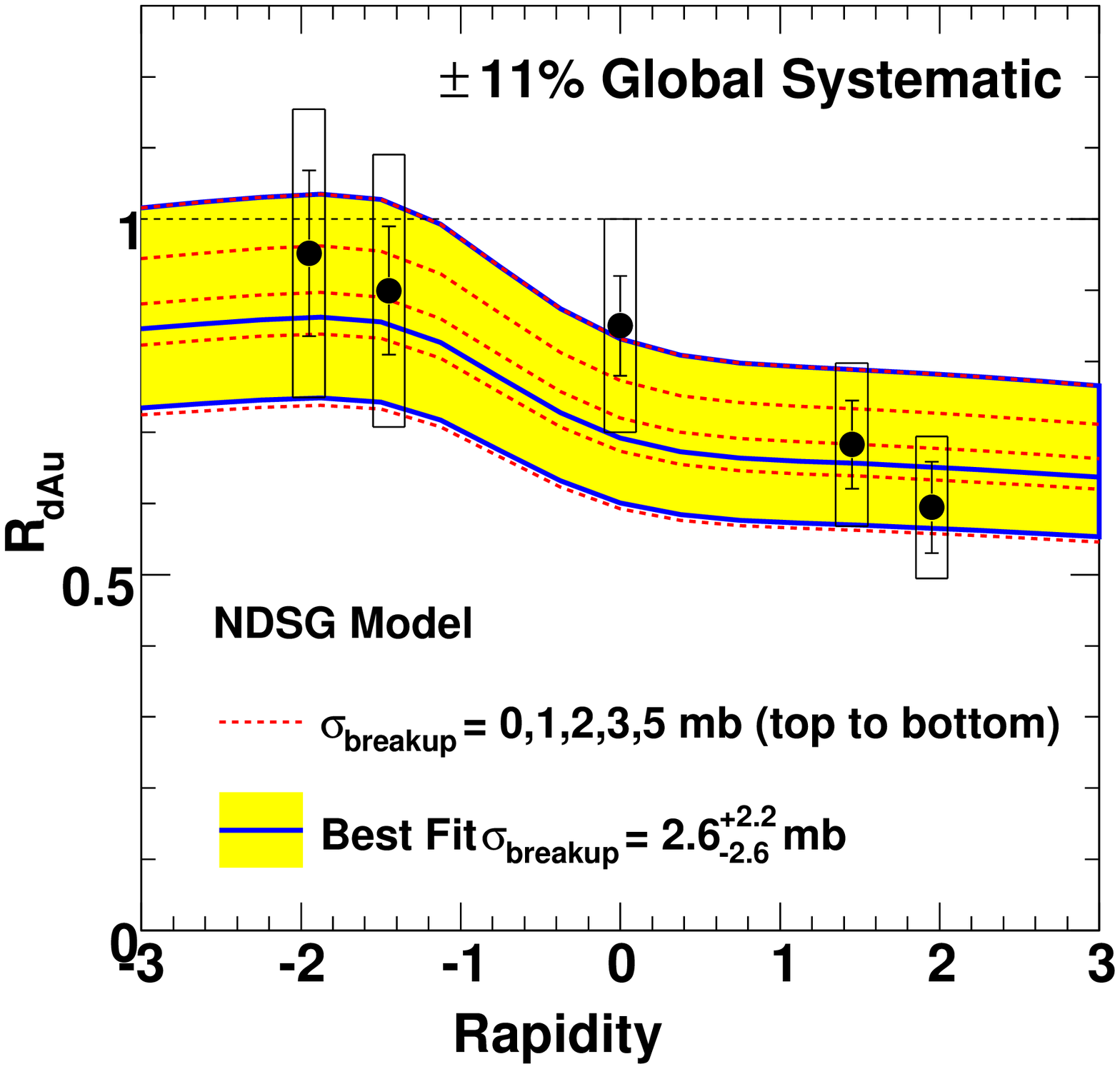}
\caption{ (color online) $\rda$ data compared to various
theoretical curves for different $\sigma_{\rm breakup}$ values.  Also,
shown as a band are the range of $\sigma_{\rm breakup}$ found to be
consistent with the data within one standard deviation.  The top panel
is a comparison for EKS shadowing~\cite{Eskola:2001ek}, while the
bottom panel is for NDSG shadowing~\cite{NDSG}.}
\label{fig:model_comparisons_dau}
\end{figure}

Thus, the one standard deviation uncertainty bands on 
$\sigma_{\rm breakup}$ in Figs.~8-11 of the original paper need 
updating with the corrected constraints.  Figures~1-4 here show 
the corrected constraints. Although the uncertainties are 
significantly larger with these corrected $\sigma_{\rm 
breakup}$ values, there is no qualitative change in the physics 
conclusions to be drawn from comparisons between these 
theoretical models (and their extrapolation) and the Au+Au and 
Cu+Cu measurements.  Stronger conclusions can only come from 
future higher statistics data for $d$+Au collisions.  
Note that no correction is needed for results from the 
data-driven method in Fig.~13 of the original paper.

\begin{table}[bth]
\centering
\caption{\label{tab:sigma_breakup} Most probable values and one
standard deviations of $\sigma_{\rm breakup}$ assuming two different
shadowing models, from a fit to minimum bias $\rda$ points as a
function of rapidity (Fig.~\ref{fig:model_comparisons_dau}), and
fits to $\rda$ as a function of $\ncol$ in three separate rapidity
bins (Fig.~\ref{fig:model_comparisons_dau_ncoll}).}
\begin{ruledtabular} \begin{tabular}{ccc}
Fit Range in $y$ & EKS (mb) & NDSG (mb) \\
\hline
All            & $2.8^{+2.3}_{-2.1}$ & $2.6^{+2.2}_{-2.6}$ \\
$[-2.2,-1.2]$  & $5.2^{+2.4}_{-2.8}$ & $3.3^{+2.9}_{-2.7}$ \\
$[-0.35,0.35]$ & $2.3^{+3.6}_{-1.9}$ & $0.8^{+3.6}_{-0.8}$ \\
$[1.2,2.2]$    & $3.4^{+2.0}_{-2.5}$ & $3.5^{+2.0}_{-2.7}$ \\
\end{tabular}
\end{ruledtabular}
\end{table}

\begin{figure}[tbh]
\includegraphics[width=1.0\linewidth]{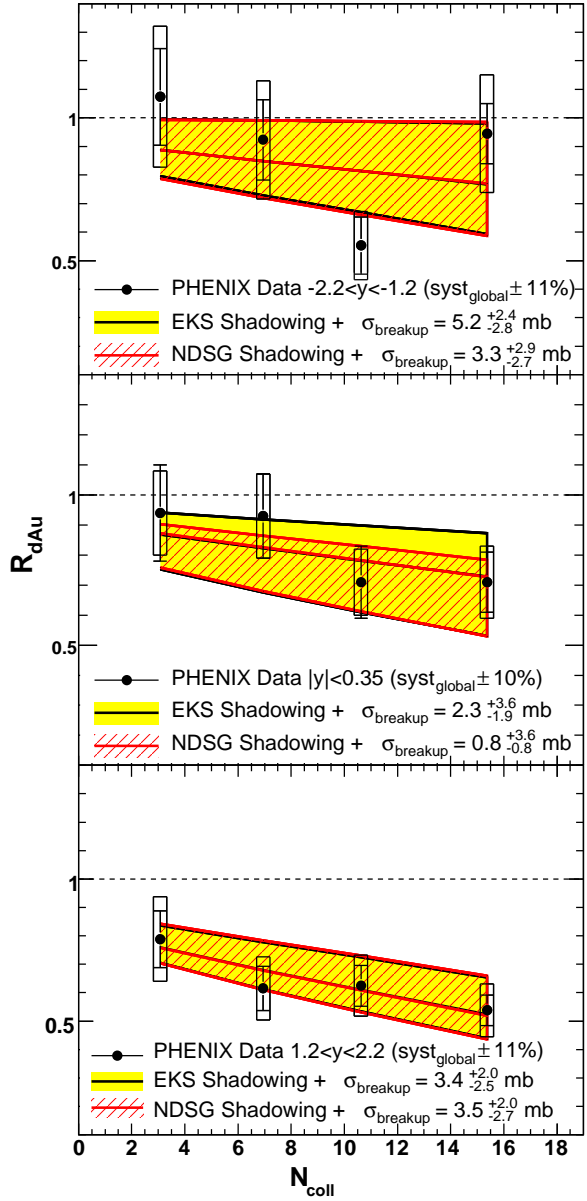}
\caption{ (color online) $\rda$ data as a function of $\ncol$ for
three different rapidity ranges.  Overlayed are theoretical curves representing
the best fit $\sigma_{\rm breakup}$ values  as determined in each rapidity range
separately, utilizing EKS and NDSG nuclear PDFs and a simple
geometric dependence.   Also, shown as bands are
the range of $\sigma_{\rm breakup}$ found to be consistent with the data
within one standard deviation.}
\label{fig:model_comparisons_dau_ncoll}
\end{figure}

\begin{figure}[tbh]
\includegraphics[width=1.0\linewidth,clip=]{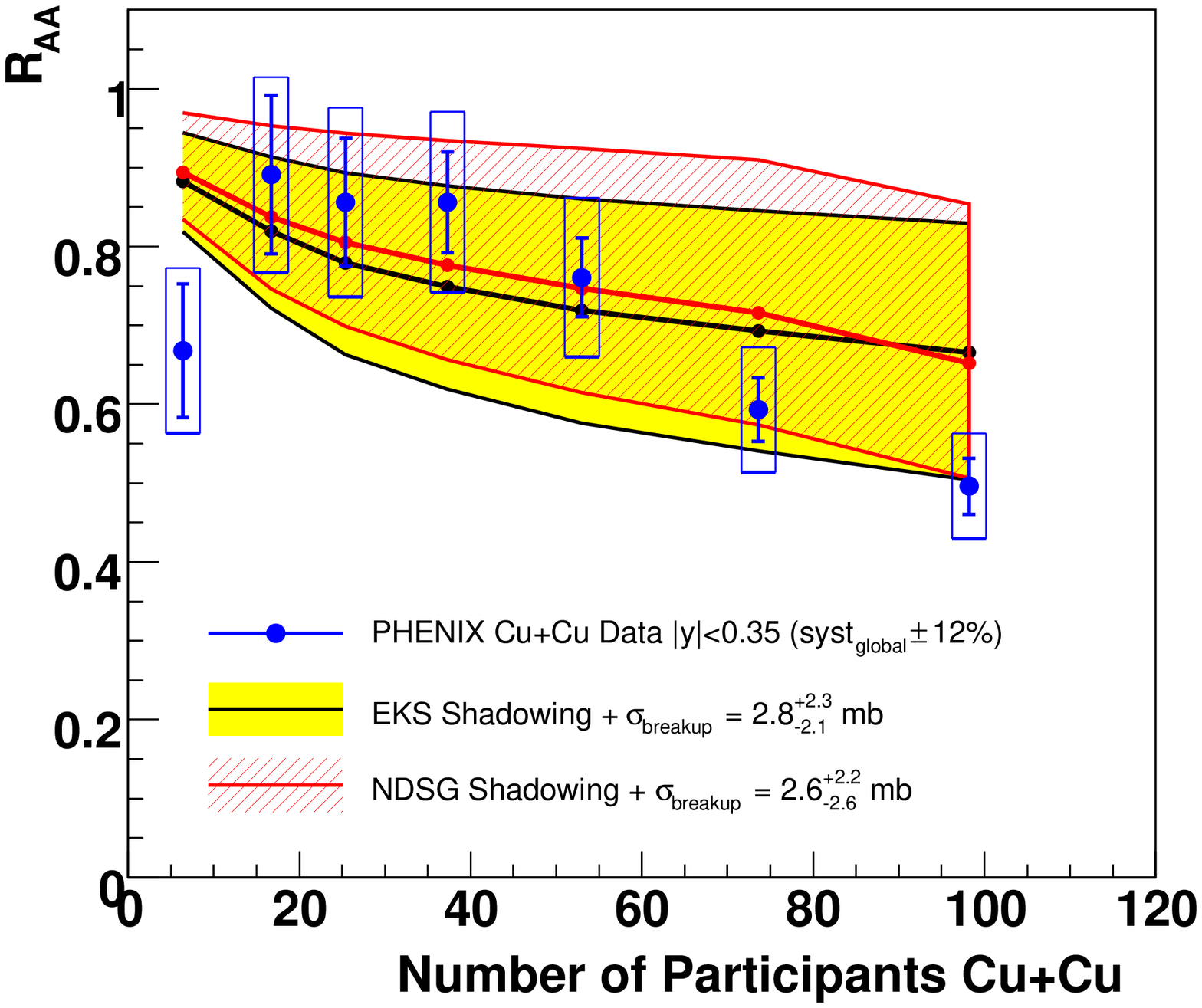}
\includegraphics[width=1.0\linewidth,clip=]{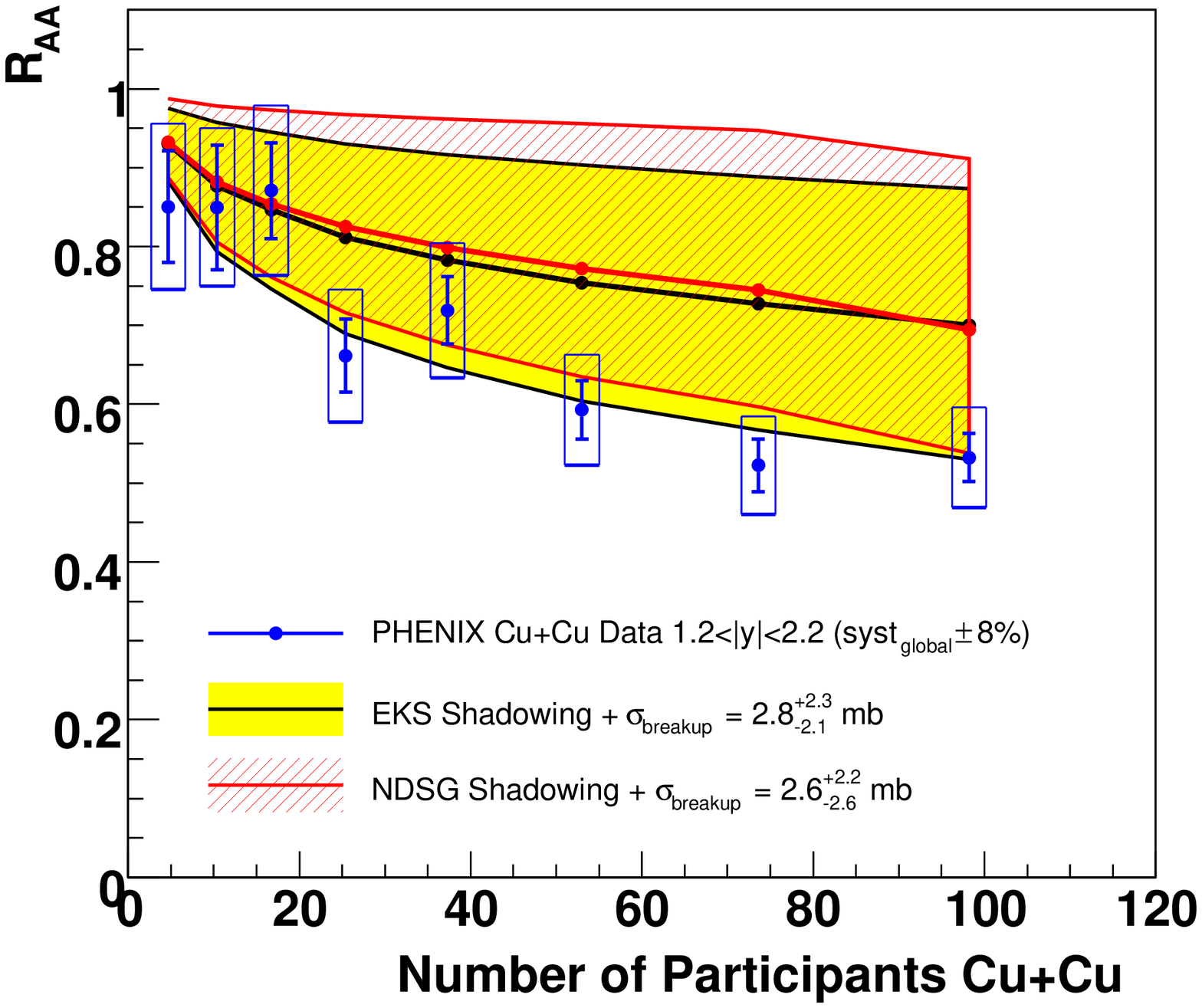}
\caption{(color online) $R_{AA}$ for Cu+Cu~\cite{Adare:2007cucu}
collisions compared to a band of theoretical curves for the
$\sigma_{\rm breakup}$ values found to be consistent with the $\dau$ data
as shown in Figure~\ref{fig:model_comparisons_dau}. The top figure
includes both EKS shadowing~\cite{Eskola:2001ek} and NDSG
shadowing~\cite{NDSG} at mid-rapidity.  The bottom figure is the same
at forward rapidity.}
\label{fig:model_comparisons_cucu} 
\end{figure}

\clearpage

\begin{figure}[tbh]
\includegraphics[width=1.0\linewidth,clip=]{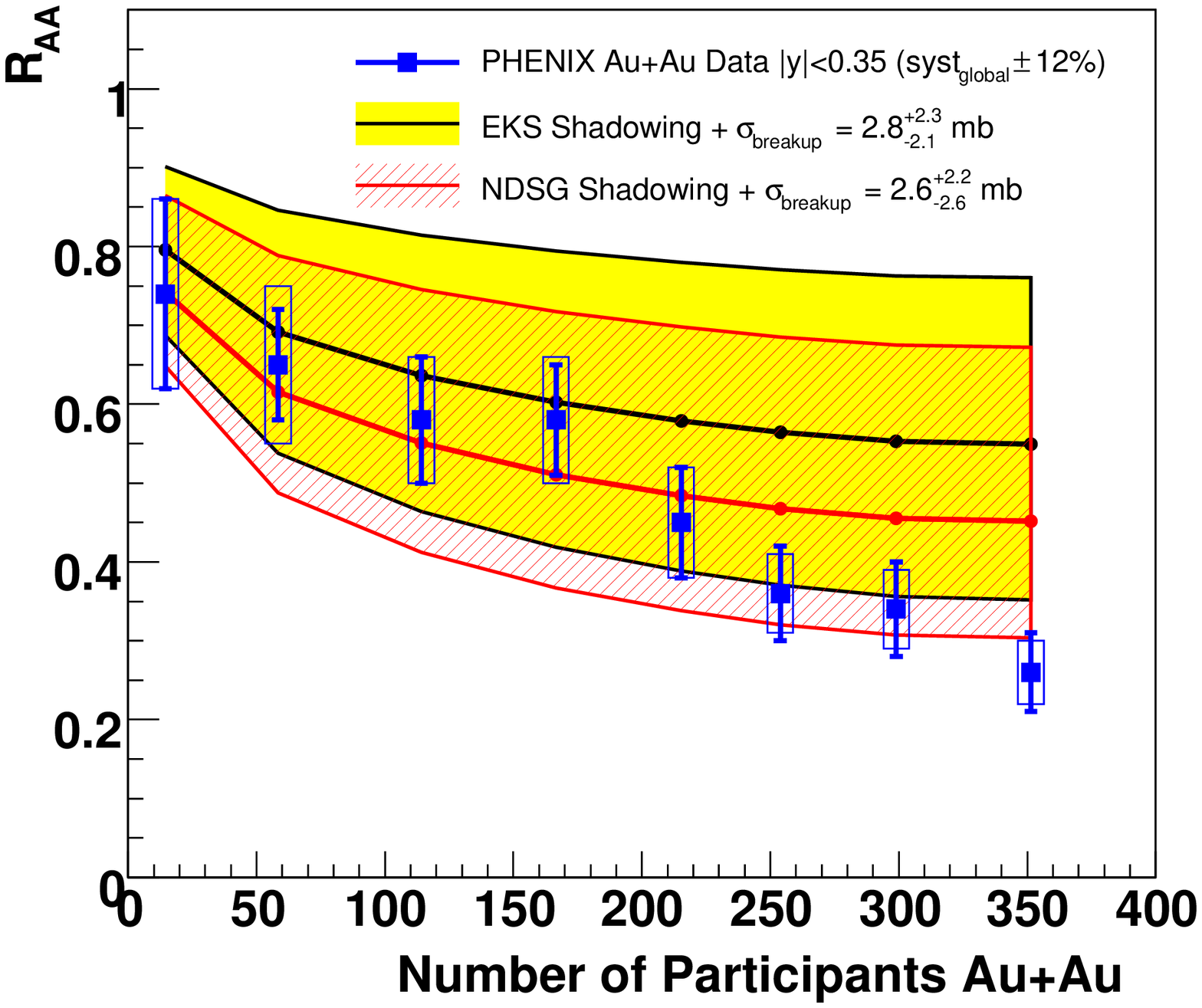}
\includegraphics[width=1.0\linewidth,clip=]{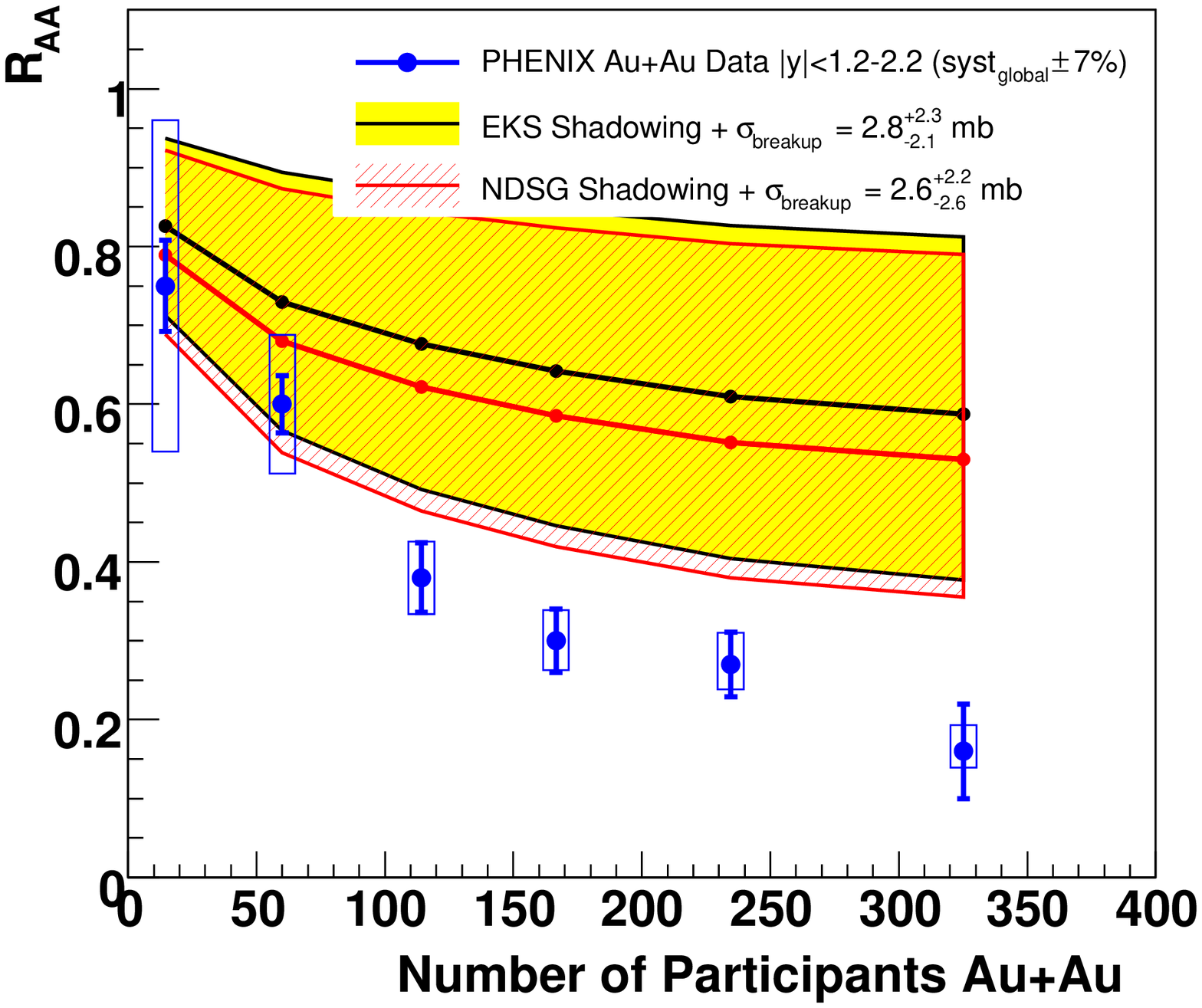}
\caption{(color online) $R_{AA}$ for Au+Au~\cite{Adare:2006ns}
collisions compared to a band of theoretical curves for the
$\sigma_{\rm breakup}$ values found to be consistent with the $\dau$ data
as shown in Figure~\ref{fig:model_comparisons_dau}. The top figure
includes both EKS shadowing~\cite{Eskola:2001ek} and NDSG
shadowing~\cite{NDSG} at mid-rapidity.  The bottom figure is the same
at forward rapidity.}
\label{fig:model_comparisons_auau} 
\end{figure}


\end{document}